\documentclass[conference]{IEEEtran}
\usepackage{cite}
\usepackage{amsmath,amssymb,amsfonts}
\usepackage{algorithmic}
\usepackage{graphicx}
\usepackage{textcomp}
\usepackage{xcolor}
\def\BibTeX{{\rm B\kern-.05em{\sc i\kern-.025em b}\kern-.08em
    T\kern-.1667em\lower.7ex\hbox{E}\kern-.125emX}}
\begin{document}

\title{Defending Against Adversarial Machine Learning}

\author{\IEEEauthorblockN{Alison Jenkins}}

\maketitle

\begin{abstract}
An Adversarial System to attack and an Authorship Attribution System (AAS) to defend itself against the attacks are analyzed. 
Defending a system against attacks from an adversarial machine learner can be done by randomly switching between models for the system, by detecting and reacting to changes in the distribution of normal inputs, or by using other methods. Adversarial machine learning is used to identify a system that is being used to map system inputs to outputs. 
Three types of machine learners are using for the model that is being attacked. The machine learners that are used to model the system being attacked are a Radial Basis Function Support Vector Machine, a Linear Support Vector Machine, and a Feedforward Neural Network. The feature masks are evolved using accuracy as the fitness measure. The system defends itself against adversarial machine learning attacks by identifying inputs that do not match the probability distribution of normal inputs. The system also defends itself against adversarial attacks by randomly switching between the feature masks being used to map system inputs to outputs. 

\end{abstract}

\begin{IEEEkeywords}
adversarial machine learning, accuracy, probability, feature mask, genetic algorithm, authorship attribution system
\end{IEEEkeywords}

\section{Introduction}

A system can defend itself against adversarial machine learners by randomly switching the feature mask it is using, by detecting unusual distributions of system inputs and reacting defensively to them, or by using other methods. Authorship attribution can be discovered using adversarial machine learning to learn the patterns between the character unigrams and authors. An adversarial machine learner can learn the mappings between system inputs and outputs and learn the Neural Network (NN) or Support Vector Machine (SVM). 

A defensive system can detect the system inputs from an adversarial machine learner by comparing the normal probability distribution of the inputs versus the current probability distribution
 of 
the inputs and detecting a difference between them. If a difference is detected between the normal inputs to the system and the current inputs to the system based on the probability, then it can be assumed that the current inputs may come from an adversarial machine learner which 
has learned 
the input-output mapping to model the NN. 
Also, if it is discovered that the inputs are along a certain decision boundary that the system has, then the defending system can become wary of the inputs and suspect them of being from an adversarial machine learning attacker system. 
Since the adversarial machine learner can develop a NN or SVM to model the system, given the input-output mappings, the adversarial machine learning system can then use knowledge of the decision boundaries in the model to inform creating inputs that will ``trick" the system that the adversarial machine learning system is attacking. 

Generative Adversarial Networks (GANs) are a type of adversarial artificial intelligence. The discriminator operates based on a neural network that has decision boundaries and makes decisions based on inputs. The generative network, the adversary, learns the distribution in order to ``fake out" the discriminator. Generative Adversarial Networks learn the distribution, create synthetic data, and can then trick the discriminator system and its decision boundaries. Examples of GANs tricking systems are ``deep fakes", where an attacker creates an input to trick the decision boundaries of the original system. The analysis of variance, such as an F-test, ANOVA test, or Student t-Test, can be done. Then, the P-value and T-values can be analyzed to make sure that the guessed value is correct. 

The seven phases in the Machine Learning Model Kill Chain are the: 
\begin{enumerate}
\item Reconnaissance (Recon) Phase
\item Weaponization Phase
\item Delivery Phase
\item Exploitation
\item Installation
\item Command and Control
\item Action
\end{enumerate} 

During the Reconnaissance (Recon) Phase, the Machine Learning (ML) models are determined. The ML models are used to protect the defending system from the type of attacks which are to be launched by the attacking system. Then, during the Weaponization Phase, the results of probes are used in an effort to develop an attack on the defending system by learning the defending system's decision boundary. During the Decision Phase, the defending system's decision boundary is attacked by the attacking adversarial machine learning system. During the Exploitation Phase, the adversarial machine learning system gathers deeper information about the defending system's underlying model. The attacking system learns how the defending system's model will be tuned, how fast new rules can be formed, and how threats are ranked. During the Installation Phase, new rules, or features, that will allow future attacks to happen, are set up. During the Command and Control Phase, a hidden command and control channel is set up to allow for expansion of the attack. Finally, during the Action Phase, the attackers act on their main objective\cite{Nyugen 2017}. 

\begin{math}30\end{math} feature masks are developed, and the system switches between the \begin{math}30\end{math} feature masks to get roughly the same results. If an attacker knows the coding language that the system is written in (Python, in this case), the random generator, and the seed, then the attacker can figure out the \begin{math}30\end{math} feature masks. 

\begin{figure}
 \begin{center}
 \setlength{\unitlength}{0.012500in}
 \includegraphics[width=40mm, scale=1.5]{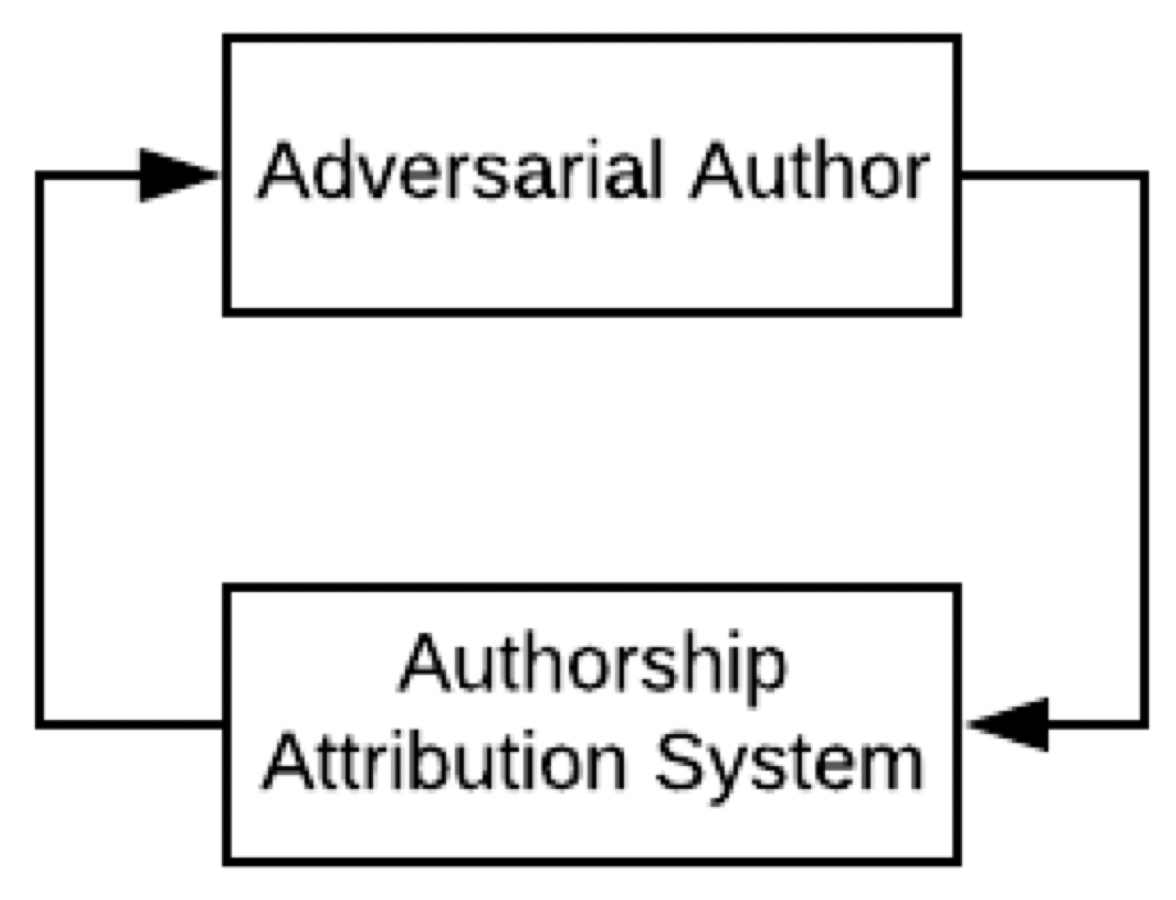}
 \end{center}
 \caption{Cycle of Adversarial Author (Attacking System) vs. Authorship Attribution System (Defending System)}
 \label{figure_AdvAuthor_AAS}
 \end{figure}

If the number of features is reduced, the accuracy will increase. Feature reduction is done \begin{math}30\end{math} times to develop the \begin{math}30\end{math} feature masks. The cycle of attack by the Adversarial Author (Attacking System) and defense by the Authorship Attribution System (AAS) (Defending System) is shown in Figure~\ref{figure_AdvAuthor_AAS}. 

Acceptable ranges of drops in accuracy by the AAS whilst under attack are less than \begin{math}11\end{math} percent. The baseline accuracy of the AAS is between \begin{math}60\end{math} to \begin{math}80\end{math} percent, depending on which algorithm is being used. Therefore, an \begin{math}11\end{math} percent drop in accuracy on \begin{math}60\end{math} to \begin{math}80\end{math} percent accuracy would result in \begin{math}49\end{math} to \begin{math}69\end{math} percent accuracy. 

Term Frequency-Inverse Document Frequency (TFIDF) is a numerical statistic that is intended to reflect how important a letter is in a character unigram. TFIDF is used as a weighting factor in searches, and the value increases proportionally to the number of times a letter appears in the character unigram and is offset by the number of character unigrams in the set that contain the letter. 

When the simple (unweighted) Euclidean distance is used, normalization provides an equal weighting to all features, whereas some features would have more importance than others when normalization is not used. Normalization of the training data is done by subtracting the mean and dividing the mean and standard deviation of the training data. 

Standardization provides a way to scale the test data with the mean and standard deviation of the training data. 

It is important for a defensive system to establish a sense of normalcy for inputs. A measure of normalcy is created by the algorithm knowing the distribution of normal inputs. If the distribution of the inputs has changed, a notification is made by the defensive algorithm to alert the system that there is something that has changed and the machine learning algorithm is receiving abnormal inputs. This violates the stationarity assumption, since the inputs are not normal and expected inputs. The inputs with abnormal distributions can trick the decision boundary and create bad results from the machine learning algorithm. 

In machine learning, Naïve Bayes classifiers are a family of simple "probabilistic classifiers" based on applying Bayes' theorem with strong (naïve) independence assumptions between the features. They are among the simplest Bayesian network models.

Naïve Bayes classifiers are highly scalable, requiring a number of parameters linear in the number of variables (features/predictors) in a learning problem. Maximum-Likelihood training can be done by evaluating a closed-form expression, which takes linear time, rather than by expensive iterative approximation as used for many other types of classifiers.

In the statistics and computer science literature, Naïve Bayes models are known under a variety of names, including simple Bayes and independent Bayes. All these names reference the use of Bayes' theorem in the classifier's decision rule, but Naïve Bayes is not (necessarily) a Bayesian method. 

The discussion so far has derived the independent feature model, that is, the Naïve Bayes probability model. The Naïve Bayes classifier combines this model with a decision rule. One common rule is to pick the hypothesis that is most probable; this is known as the Maximum A-Posteriori (MAP) decision rule. 

When dealing with continuous data, a typical assumption is that the continuous values associated with each class are distributed according to a normal (or Gaussian) distribution. For example, suppose the training data contains a continuous attribute, \begin{math}x\end{math}. The data is first segmented by the class, and then compute the mean and variance of \begin{math}x\end{math} in each class.

The algorithms are sorted into equivalence classes based on whether there are statistically significant differences between the algorithms. The ANOVA and Student t-Tests are used to determine statistical significance and to compare the algorithms.

%
%

\section{Methodology}

An Authorship Attribution System (AAS) that is resistant to Adversarial Authorship Attacks is developed. After developing the AAS, a method for testing it is developed. The AAS reads from a file that contains the adversarial texts, called ``AdversarialTests.txt". This file contains the list of adversarial texts that will be located in the same directory as the AAS. The ``AdversarialTests" will contain the file names of a number of adversarial texts (advText00.txt, advText01.txt, ..., advText24.txt). The AAS reads in each adversarial text and classifies it by placing the results in a file, called ``AdversarialTestResults.txt". The ``AdversarialTestResults.txt" file will have a classification associated with each adversarial text of the form: 
\begin{itemize}
\item advText00.txt \begin{math}\rightarrow 1000\end{math}
\item advText01.txt \begin{math}\rightarrow 1022\end{math}
\item \begin{math}...\end{math}
\item advText24.txt \begin{math}\rightarrow 1005\end{math}
\end{itemize}

\subsection{Defending System}
Principle Component Analysis (PCA) may be used to reduce the number of features need for the Feature Mask to correctly identify the author. Principal Component Analysis consists of keeping the best and most effective (or highest magnitude) features and then either not using or getting rid of the other features. np.count\_nonzero() is a function that can be used to count the number of features that are being used in the feature mask. The reduction of features results in an increase in accuracy. 

Switching between Feature Masks may be used as a way to keep the attacking system from learning the decision boundaries of the defending system. Using this method, the defending system randomly switches between equivalent feature masks. 

Use a vector of attack criteria to decide whether or not the defending system thinks that it is being attacked. If multiple factors give positive values, then the system will determine that it must currently be under attack. Then, the defending system will respond accordingly to the perceived attack. 

\subsection{Attacking System}
The defending system may be attacked during any of the \begin{math}5\end{math} Stages of Machine Learning. The \begin{math}5\end{math} Stages of Machine Learning are: 
\begin{enumerate}
\item Measuring
\item Feature Selection
\item Learning Model 
\subitem - Lazy Learner 
\subsubitem * General Regression Neural Network (GRNN) 
\subsubitem * No Model
\subitem - Support Vector Machine (SVM)
\subitem - Feedforward Neural Network (FFNN))
\item Training
\item Prediction
\end{enumerate}

There is a conditional probability, or an association, between the inputs the defending system should expect to receive and the inputs that it is actually receiving. If the system begins to receive inputs that do not match the distribution that it is expecting, then the system can recognize that the input distribution is different and react accordingly. This makes the attacking system have to then change its method of attack, in order to continue attacking the defending system in an effective manner. 

The effect of feedback control on measurement noise is that the system handles measurement noise well, if the noise is Gaussian white noise. 
Therefore, the system may give misleading outputs. One possibility to avoid mistakes is to compute the correlation of the equation error and check whether it is white noise. The decision boundaries are ``tricked" or ``fooled" by modifying the input based on knowing how the system will respond to changes in the input. This is a technique used by adversarial machine learning attackers to trick a system's decision boundaries. The defending system must find a way either to mitigate the effect of, or to recognize, the attack. 

The Steady-State Genetic Algorithm (SSGA) is used to evolve the feature masks in a way that increases the accuracy. The SSGA replaces the worst fit individual in the population with the child. A population size of \begin{math}30\end{math} is used. A mutation rate of \begin{math}0.5\end{math} is used. There are \begin{math}95\end{math} features in each feature vector, and there are \begin{math}30\end{math} feature masks. To clarify, each feature mask is one member of the population, so there are \begin{math}30\end{math} feature masks, or population members. Binary Tournament Selection is used for selecting parents from the feature masks. Uniform Crossover is used to produce the child's features from the parents' features. 
%
%
%
%
%
%
%
%
\subsection{Radial Basis Function Support Vector Machine}

The Radial Basis Function Support Vector Machine (RBFSVM) is implemented using scikit-learn with its default kernel being a radial basis function and this is a method whose value depends on the distance from the origin or from some point. In this paper, the output received from the processing model is then evaluated on the RBFSVM model and the accuracy is obtained.

\subsection{Linear Support Vector Machine}

The Linear Support Vector Machine (LSVM) is implemented using scikit-learn. The LSVM is the simplest form of a support vector machine. Along with the RBFSVM, the output of the processing model is passed through the LSVM and its accuracy is also obtained.
%
%
%
%
%
%
%
%
%

\subsection{Feedforward Neural Network}


The advantages of Multi-Layer Perception (MLP) are its capability to learn nonlinear models and its capability to learn models in real-time (online learning). 
There are three main disadvantages of MLP. The first disadvantage is that MLP with hidden layers has a non-convex loss function where there exists more than one local minimum. Therefore, different random weight initializations can lead to different validation accuracy. Secondly, MLP requires tuning a number of hyper-parameters such as the number of hidden neurons, layers, and iterations. Lastly, MLP is sensitive to feature scaling. 

NNs may be very large, so it is impractical to write down the gradient formula by hand for all parameters. Back-propagation is the recursive application of the Chain Rule along a computational graph to compute the gradients of all inputs/parameters/intermediates. The implementations of back-propagation maintain a graph structure, where the nodes implement forward()/backward() API. The forward pass computes the result of an operation and saves any intermediates needed for gradient computation in memory. The backward pass applies the Chain Rule to compute the gradient of the loss function with respect to the inputs. 

\section{Experiment}

Each author uses \begin{math}4\end{math} writing samples, and there are \begin{math}25\end{math} authors. There are thus \begin{math}100\end{math} training sets. There are \begin{math}7\end{math} letters in the alphabet which are used. A Genetic Algorithm (GA) with uniform crossover is used.
The algorithm is run \begin{math}30\end{math} times, and the best feature mask is found out of all \begin{math}30\end{math} of the runs. 
The best fit individual is extracted from the character unigrams, or the number of features. 
The feature consistency is the consistency of a feature being present over \begin{math}30\end{math} runs.
The feature mask consists of \begin{math}1\end{math}'s and \begin{math}0\end{math}'s. 
The feature mask multiplied by the learning set vector results in the derived learning set vector gets rid of the components that are not to be considered. 

For feature selection, a population is randomly generated. The population size that is used is \begin{math}30\end{math}. The mutation rate is \begin{math}0.5\end{math}. There are \begin{math}100\end{math} feature vectors. The feature extraction is used to come up with the dataset, or feature set. 
The fitness value is found for each feature. In each algorithm, accuracy is used for the fitness value. In other words, the accuracy for the derived training set is used for determining the fitness for each individual in the population. 


The evolutionary process is continued until a stopping rule is reached. Stopping rules that can be used are number of function evaluations, reaching a certain fitness value, or another stopping criteria may be used. The algorithm used in the test had a stopping rule of the limiting number of function evaluations being \begin{math}1000\end{math} and the fitness value, or accuracy, reaching \begin{math}0.99754\end{math}. This means that the algorithm must run until it reaches either the limit of \begin{math}1000\end{math} function evaluations or \begin{math}99.754\%\end{math} accuracy, depending on whichever stopping limit is reached first.

\begin{table}[ht]
\caption{Algorithm Dataset Allocation}
\def\arraystretch{1.1}
\begin{center}
\begin{tabular}{|c|c|c|c|c|}
\hline
\textbf{Algorithm} & \textbf{Training} & \textbf{Evaluation} & \textbf{Validation} & \textbf{Test} \\
\hline
\textbf{\textit{GRNN}} & 70\% & 10\% & 10\% & 10\% \\
\textbf{\textit{RBFNN}} & 80\% & 0\% & 10\% & 10\% \\
\textbf{\textit{SVM}} & 90\% & 0\% & 0\% & 10\% \\
\textbf{\textit{FFNN}} & 80\% & 0\% & 10\% & 10\% \\
\hline
\end{tabular}
\label{table:Algorithm_Dataset Allocation_Summary}
\end{center}
\end{table}

\subsection{Radial Basis Function Support Vector Machine}

For the RBFSVM model, the test set is evaluated for the default values of the RBFSVM function available built into the function in scikit-learn. The function is passed with train data and train labels as parameter for training purpose. It is evaluated with the evaluation set and evaluation labels as parameters which are then used to obtain the accuracy.

\subsection{Linear Support Vector Machine}

For the LSVM model, the linear kernel is used to process the evaluation set. The function is passed with train data and train labels as parameter for training purpose. It is then evaluated with evaluation set and evaluation labels as parameters which are used to obtain the accuracy.

\subsection{Feedforward Neural Network}

The MultiLayer Perceptron (MLP) function is used to train the model with training data and train labels for the FFNN and then the evaluation data and evaluation labels are used to test the model for accuracy.

\section{Results}

The ANOVA Tests gives a P-value \begin{math}{< 0.5}\end{math} which suggests that there is no statistical significant difference and has weak acceptance of hypothesis. This means that all the three models are in different equivalence classes. The Student t-Test values for all combinations of RBFSVMs, LSVMs and FFNNs have t-Stat value \begin{math}{< -0.5}\end{math} which suggests that the null hypothesis is accepted for each combination. The results of the statistical analysis of a comparison of all the algorithms is shown in Table~\ref{table_comparison_results}. 

\begin{table}[ht]
\caption{Comparison of Algorithms}
\def\arraystretch{1.1}
\begin{center}
\begin{tabular}{|c|c|c|c|}
\hline
\textbf{}&\textbf{}&\textbf{}&\textbf{} \\
\textbf{Run} & \textbf{\textit{RBFSVM}} & \textbf{\textit{LSVM}} & \textbf{\textit{FFNN}} \\
\hline
1 & 0.68 & 0.72 & 0.68 \\
2 & 0.72 & 0.64 & 0.68 \\
3 & 0.6 & 0.72 & 0.76 \\
4 & 0.68 & 0.72 & 0.68 \\
5 & 0.68 & 0.72 & 0.72 \\
6 & 0.6 & 0.68 & 0.68 \\
7 & 0.68 & 0.72 & 0.72 \\
8 & 0.56 & 0.68 & 0.68 \\
9 & 0.6 & 0.68 & 0.72 \\
10 & 0.56 & 0.72 & 0.76 \\
11 & 0.68 & 0.72 & 0.84 \\
12 & 0.6 & 0.8 & 0.72 \\
13 & 0.6 & 0.72 & 0.76 \\
14 & 0.6 & 0.68 & 0.8 \\
15 & 0.6 & 0.76 & 0.76 \\
16 & 0.64 & 0.68 & 0.76 \\
17 & 0.64 & 0.68 & 0.68 \\
18 & 0.64 & 0.68 & 0.72 \\
19 & 0.6 & 0.72 & 0.72 \\
20 & 0.6 & 0.84 & 0.72 \\
21 & 0.6 & 0.68 & 0.68 \\
22 & 0.6 & 0.72 & 0.68 \\
23 & 0.68 & 0.64 & 0.76 \\
24 & 0.68 & 0.8 & 0.76 \\
25 & 0.64 & 0.76 & 0.76 \\
26 & 0.6 & 0.68 & 0.64 \\
27 & 0.72 & 0.76 & 0.76 \\
28 & 0.6 & 0.68 & 0.76 \\
29 & 0.64 & 0.8 & 0.68 \\
30 & 0.68 & 0.76 & 0.72  \\
\hline
\textbf{\textit{Average}} & 0.633 & 0.719 & 0.725 \\
\hline
\end{tabular}
\label{table_comparison_results}
\end{center}
\end{table}

The results place 
the three algorithms in the same equivalence class using both the ANOVA and Student t-Tests. When the ANOVA test and the Student t-Tests are performed, the ANOVA test of 
the three algorithms yields a p-value of 3.35E-12, so the F-Test is then performed to determine which two-tailed two-sample Student t-Test to use. In each comparison between algorithms, the Student t-Test results in a t-Stat value that is smaller than the t Critical value. Therefore, the null hypothesis is accepted.

The results from each algorithm are shown in Table~\ref{table_comparison_results}, while the ANOVA test results are shown in Tables~\ref{table:ANOVA_Single_Factor_Summary} and~\ref{table:ANOVA_Single_Factor_Var_Summary}.

\begin{table}[ht]
\caption{ANOVA Test Summary}
\def\arraystretch{1.1}
\begin{center}
\begin{tabular}{|c|c|c|c|c|}
\hline
\textbf{}&\textbf{}&\textbf{}&\textbf{}&\textbf{} \\
\textbf{Groups} & \textbf{Count} & \textbf{Sum} & \textbf{Average} & \textbf{Variance} \\
\hline
\textbf{\textit{RBFSVM}} & 30 & 19.08 & 0.636 & 0.00180 \\
\textbf{\textit{LSVM}} & 30 & 21.56 & 0.719 & 0.00237 \\
\textbf{\textit{FFNN}} & 30 & 21.76 & 0.725 & 0.00196 \\
\hline
\end{tabular}
\label{table:ANOVA_Single_Factor_Summary}
\end{center}
\end{table}

\begin{table}[ht]
\caption{ANOVA Test Variation Summary}
\def\arraystretch{1.1}
\begin{center}
\begin{tabular}{|c|c|c|c|c|c|c|}
\hline
\textbf{}&\textbf{}&\textbf{}&\textbf{}&\textbf{}&\textbf{}&\textbf{} \\
\textbf{\begin{tabular}{@{}c@{}}Source of\\Variation\end{tabular}}&\textbf{SS}&\textbf{df}&\textbf{MS}&\textbf{F}&\textbf{P-value}&\textbf{F crit}\\
\hline
\textbf{\textit{Between}}& 0.149 & 2 & 0.074 & 36.35 & 3.35E-12 & 3.10 \\
\textbf{\textit{Within}}& 0.178 & 87 & 0.002 &  &  &  \\
\textbf{\textit{Total}}& 0.326 & 89 &  &  &  &  \\
\hline
\end{tabular}
\label{table:ANOVA_Single_Factor_Var_Summary}
\end{center}
\end{table}

Representative Student t-Tests are shown in Tables~\ref{table:tTest_RS_SS} and~\ref{table:tTest_SS_SA}.

\begin{table}[ht]
\caption{Student t-Test: Two-Sample Assuming Equal Variances}
\def\arraystretch{1.1}
\begin{center}
\begin{tabular}{|c|cc|}
\hline
\textbf{}&\textbf{}&\textbf{} \\
\textbf{} & \textbf{RBFSVM} & \textbf{LSVM} \\
\hline
\textbf{Mean} & 0.636 & 0.718 \\
\hline
\textbf{Variance} & 0.00180 & 0.00237 \\
\hline
\textbf{Observations} & 30 & 30 \\
\hline
\textbf{Pooled Variance} & 0.00209 & \\
\hline
\textbf{Hypothesized Mean Difference} & 0 & \\
\hline
\textbf{df} & 58 & \\
\hline
\textbf{t-Stat} & -7.008 & \\
\hline
\textbf{P(T\begin{math}{<=}\end{math}t) one-tail} & 1.42E-09 & \\
\hline
\textbf{t Critical one-tail} & 1.671 & \\
\hline
\textbf{P(T\begin{math}{<=}\end{math}t) two-tail} & 2.84E-09 & \\
\hline
\textbf{t Critical two-tail} & 2.002 & \\
\hline
\end{tabular}
\label{table:tTest_RS_SS}
\end{center}
\end{table}

\begin{table}[ht]
\caption{Student t-Test: Two-Sample Assuming Unequal Variances}
\def\arraystretch{1.1}
\begin{center}
\begin{tabular}{|c|cc|}
\hline
\textbf{}&\textbf{}&\textbf{} \\
\textbf{} & \textbf{LSVM} & \textbf{FFNN} \\
\hline
\textbf{Mean} & 0.719 & 0.725 \\
\hline
\textbf{Variance} & 0.00237 & 0.00196 \\
\hline
\textbf{Observations} & 30 & 30 \\
\hline
\textbf{Hypothesized Mean Difference} & 0 & \\
\hline
\textbf{df} & 57 &  \\
\hline
\textbf{t-Stat} & -0.555 & \\
\hline
\textbf{P(T\begin{math}{<=}\end{math}t) one-tail} & 0.290 & \\
\hline
\textbf{t Critical one-tail} & 1.672 & \\
\hline
\textbf{P(T\begin{math}{<=}\end{math}t) two-tail} & 0.581 & \\
\hline
\textbf{t Critical two-tail} & 2.002 & \\
\hline
\end{tabular}
\label{table:tTest_SS_SA}
\end{center}
\end{table}

There are \begin{math}2\end{math} equivalence classes. The LSVM with Normalization, Standardization, and TFIDF and the FFNN with Normalization, Standardization, and TFIDF are in the same equivalence class, while the RBSVM with Normalization, Standardization, and TFIDF is in a different equivalence class.

The resulting accuracies are shown in Table~\ref{table_variations_comparison_results}. 

 \begin{table}[htbp]
 \caption{Effects of Normalization, Standardization, and TFIDF on Algorithm Accuracy}
 \begin{center}
 \begin{tabular}{|c|c|c|c|}
 \hline
 \cline{2-4} 
 \textbf{Algorithm} & \textbf{\textit{RBFSVM}} & \textbf{\textit{LSVM}} & \textbf{\textit{FFNN}} \\
 \hline
 None & 0.54 &  0.32 &  0.42 \\
 Normalization, Standardization, TFIDF & 0.7 &  0.62 &  0.682 \\
 Standardization, TFIDF & 0.51 &  0.52 &  0.61 \\
 Normalization, TFIDF & 0.46 &  0.48 &  0.601 \\
 Normalization, Standardization & 0.7 &  062 &  0.674 \\
 Normalization & 0.59 &  0.29 &  0.44 \\
 Standardization & 0.51 &  0.52 &  0.65 \\
 TFIDF & 0.42 &  0.47 &  0.6 \\
 \hline
 \end{tabular}
 \label{table_variations_comparison_results}
 \end{center}
 \end{table}

The results of the statistical analysis of a comparison of all the algorithms is shown in Table~\ref{table_comparison_results}.

 \begin{table}[htbp]
 \caption{Comparison Results}
 \begin{center}
 \begin{tabular}{|c|c|c|c|c|}
 \hline
 \textbf{}&\multicolumn{4}{|c|}{\textbf{}} \\
 \cline{2-5} 
 \textbf{Algorithm} & \textbf{\textit{Accuracy}} & \textbf{\textit{Precision}} & \textbf{\textit{Recall}} & \textbf{\textit{F1}} \\
 \hline
 GRNN & 0.4315 &  0.3693 &  0.2457 &  0.2571 \\
 EGRNN & 0.4328 &  0.4163 &  0.3682 &  0.3369 \\
 RBFNN$^{\mathrm{1}}$& 0.2367 &  0.3454 &  0.2335 &  0.2342 \\
 RBFNN$^{\mathrm{2}}$& 0.2735 &  0.4245 &  0.3253 &  0.3726 \\
 SVM$^{\mathrm{3}}$ & 0.4342 &  0.4137 &  0.4577 &  0.4643 \\
 SVM$^{\mathrm{4}}$ & 0.3172 &  0.2845 &  0.3435 &  0.2281 \\
 FFNN & 0.1293 &  0.1346 &  0.3946 &  0.2986 \\
 \hline
 \multicolumn{5}{l}{$^{\mathrm{1}}$Without Kohonen Unsupervised Learning and Back-Propagation} \\
 \multicolumn{5}{l}{$^{\mathrm{2}}$With Kohonen Unsupervised Learning and Back-Propagation} \\
 \multicolumn{5}{l}{$^{\mathrm{3}}$Linear SVM} \\
 \multicolumn{5}{l}{$^{\mathrm{4}}$Radial Basis SVM} \\
 \end{tabular}
 \label{table_comparison_results}
 \end{center}
 \end{table}

%
%

%
%
%

\subsection{Radial Basis Function Support Vector Machine}

The effect of not using TFIDF is minimal. The accuracies produced are, on average, exactly the same as the accuracies using the TFIDF.
The effect of not using Standardization is an average drop in accuracy of about 10\%. If a dataset is not standardized before it is used and does not naturally have a normal distribution, then the algorithm is going to perform badly. 
The effect of not using Normalization is not a large change in accuracy values. The average for 5 runs is approximately the same with and without normalization. Since the accuracies cannot be negative values, the effect of normalization on all three of the algorithms is minimal. If the dataset produced negative values, the effect would be more noticeable. 

\subsection{Linear Support Vector Machine}

The effect of not using TFIDF is basically no change to the accuracy values. The accuracies are within the same range when TFIDF is used or is not used. 
The effect of not using Standardization is also an average drop in accuracy of about 10\%. The StandardScalar in scikit-learn removes the mean and scales the data to unit variance. This still allows for outliers to influence the computation, and therefore does not guarantee balanced feature scales in the cases of having outliers.
The effect of not using Normalization is a drop in accuracy of about 10\%. The LSVM is the most affected by not using normalized data. 

\subsection{Feedforward Neural Network}

The effect of not using TFIDF also results in no change to the accuracy values. 
The effect of not using Standardization is a convergence warning. The FFNN has a maximum iteration value of 2000, and when the data is not standardized, the optimization cannot converge in the 2000 iterations.
The effect of not using Normalization is minimal. It results in a drop of about 5\%, at the most. 

\section{Conclusions}
%

The RBFSVM, LSVM, and FFNN are all affected in similar ways when TFIDF, Standardization, and Normalization are not used. 
It is evident from the ANOVA test and Student t-Test that the LSVM and FFNN are in the same equivalence class. The FFNN had the highest average accuracy for 30 function evaluations, with an average accuracy of 72.5\%. The ANOVA Test gives a P-value \begin{math}{< 0.05}\end{math}, which both suggests that there is no statistical significant difference and has weak acceptance of the hypothesis. This means that all the three models are in different equivalence classes. The Student t-Test values for all combinations of RBFSVMs, LSVMs and FFNNs have t-Stat value \begin{math}{< -0.05}\end{math} which suggests that the null hypothesis is accepted for each combination.

Particle Swarm Optimization (PSO) is an option for an algorithm that may be used instead a GA and may be explored in future work.

\section{Breakdown of the Work}

Alison Jenkins - Research, Code, Analysis, {\LaTeX} Report, and Extra Credit.

\vspace{12pt}
\color{red}

\end{document}